\documentclass[a4paper,british]{extarticle}
\usepackage[T1]{fontenc}
\usepackage[utf8x]{inputenc}
\usepackage{babel}
\usepackage{booktabs}
\usepackage{float}
\usepackage{subfig}
\usepackage{authblk}
\usepackage{mathtools}
\usepackage{amsmath}
\usepackage{amsthm}
\usepackage{amssymb}
\usepackage{graphicx}
\usepackage{nomencl}
\usepackage[
 unicode=true,
 bookmarks=true,
 bookmarksnumbered=false,
 bookmarksopen=false,
 breaklinks=true,
 pdfborder={0 0 0},
 pdfborderstyle={},
 backref=false,
 colorlinks=false
]{hyperref}
\usepackage[capitalise]{cleveref}

\DeclareMathOperator{\EX}{\mathbb{E}} 
\DeclareMathOperator\var{Var}         

\makeatletter

\newtheorem{definition}{Definition}
\newtheorem{lemma}{Lemma}

\@ifundefined{showcaptionsetup}{}{%
 \PassOptionsToPackage{caption=false}{subfig}}
\makeatother

\providecommand{\keywords}[1]
{
  \small	
  \textbf{\textit{Keywords---}} #1
}

\makenomenclature

\renewcommand{\nompreamble}{Throughout the paper, boldface is used for vectors and capital boldface for matrices: to provide some examples, the vector of RGD, the target variable, is denoted by $\mathbf{y}$, the matrix of the inputs is $\mathbf{X}$ and its $i$-th column is $\mathbf{x}_i$. \\ The main symbols and the notation adopted in the paper are summarised below.}

\begin{document}

\title{Forecasting residential gas demand: machine learning approaches and seasonal role of temperature forecasts}

\author[1]{Emanuele Fabbiani}
\author[1]{Andrea Marziali}
\author[1]{Giuseppe De Nicolao}
\affil[1]{Department of Electrical, Computer and Biomedical Engineering, University of Pavia}

\maketitle

\begin{abstract}
Gas demand forecasting is a critical task for energy providers as it impacts on pipe reservation and stock planning. In this paper, the one-day-ahead forecasting of residential gas demand at country level is investigated by implementing and comparing five models: Ridge Regression, Gaussian Process (GP), k-Nearest Neighbour, Artificial Neural Network (ANN), and Torus Model. Italian demand data from 2007 to 2017 are used for training and testing the proposed algorithms. The choice of the relevant covariates and the most significant aspects of the pre-processing and feature extraction steps are discussed in depth, lending particular attention to the role of one-day-ahead temperature forecasts. Our best model, in terms of Root Mean Squared Error (RMSE), is the ANN, closely followed by the GP. If the Mean Absolute Error (MAE) is taken as an error measure, the GP becomes the
best model, although by a narrow margin. A main novel contribution is the development of a model describing the propagation of temperature errors to gas forecasting errors that is successfully validated on experimental data. Being able to predict the quantitative impact of temperature forecasts on gas forecasts could be useful in order to assess potential improvement margins associated with more sophisticated weather forecasts. On the Italian data, it is shown that temperature forecast errors account for some 18\% of the mean squared error of gas demand forecasts provided by ANN.

\end{abstract} 
\hspace{10pt}

\keywords{natural gas; time series forecasting; statistical learning; Gaussian Process; neural networks.}

\newpage{}

\section{Introduction}

Forecasting natural gas demand is a crucial task for energy companies for several reasons. First, it provides relevant information to reserve pipe capacity and plan stocks effectively. Furthermore, regulations impose the balance of the network by charging providers with a fee proportional to their unbalanced quantity. Finally, demand is a critical input to forecast gas price, which is, in turn, a driver for business decisions.

Two comprehensive reviews of the literature about gas demand forecasting are \cite{vsebalj2017predicting} and \cite{soldo2012forecasting}. According to \cite{vsebalj2017predicting}, papers can be classified along four dimensions. The \textit{prediction horizon} can range from hourly to yearly, the \textit{reference area} from single nodes of the network to a whole country; adopted \textit{models} include time series, mathematical and statistical approaches, neural networks, and others; input \textit{features} can be demand history, temperature, calendar, and other minor ones. 

Several studies focused on country- or regional-level daily forecasting. Mathematical and statistical models based on non-linear parametric functions were used in \cite{brabec2008nonlinear} to explain the factors which affect the demand.
A different multi-factor approach was developed in \cite{potovcnik2007forecasting} and a model based on the physical relation between gas demand and the temperature was presented in \cite{gil2004generalized}. An adaptive-network-based fuzzy inference system (ANFIS) was described in \cite{azadeh2010adaptive}, where the authors showed better performances of their model concerning ANN and conventional time series methods. A statistical learning model, based on support vector machine (SVM), was developed in \cite{zhu2015short} for UK demand, and compared to ANN and an autoregressive moving average (ARMA) predictor. A hybrid model, exploiting many different techniques, such as wavelet transform, genetic algorithm, ANFIS, and ANN, was used in \cite{panapakidis2017day}. Neural networks were applied in \cite{tacspinar2013forecasting, demirel2012forecasting,szoplik2015forecasting,soldo2014improving,tonkovic2009predicting} to perform hourly and daily forecasts on cities and regions. Moreover, \cite{wei2019daily} showed how ANNs, combined with Principal Components Correlation Analysis (PCCA), provide robust and precise forecasts on regional demand. Baldacci et al. \cite{BALDACCI2016190} used nearest neighbors and local regression to forecast the gas demand of small villages. They also presented an investigation over the effects of temperature forecast errors, concluding that the influence on model accuracy is negligible.

Concerning long-term forecasting, \cite{Wadud2011Modeling} discussed gas demand in Bangladesh, showing how population growth and Gross Domestic Product (GDP) are essential drivers of the demand. Similar conclusions were achieved in \cite{Karadede2017Breeder}, where a breeder model was proven superior to other approaches in forecasting Turkish demand.

The present work focuses on day-ahead forecasting of Residential Gas Demand (RGD) at country level. A companion paper focuses on Industrial (IGD) and Thermoelectric Gas Demand (TGD) \cite{marziali2019ensembling}. The focus of the analysis is on two issues, not adequately covered in the existing literature, with Italian RGD used as a case study. 

The first issue addressed is the comparison between regression and machine learning methods in order to understand what technique is more suitable when performance is assessed over several years. Five forecasting methods are considered, two based on linear regression and three on machine learning techniques, paying particular attention to their accurate tuning. This involves a detailed discussion on the selection of the relevant covariates, among which a primary role is played by the weather temperature.

The second issue has in fact to do with the influence of weather forecast errors on natural gas demand models. Despite being critical in industrial applications, previous works seldom specify if the predictors use forecasted or observed temperature, maybe due to the belief that temperature errors have negligible impact. In contrast, we assess the influence of weather forecasting errors, both theoretically and experimentally. A novel easy-to-compute bound is derived that predicts the best achievable RGD root mean square error (RMSE) as a function of the temperature RMSE. This bound is then validated on experimental data: Italian RGD forecasts are obtained using both observed and predicted temperatures, thus allowing for a quantitative assessment of accuracy degradation.

The paper is organized as follows. In \cref{Problem Statement}, we formulate the problem and present the available data. In \cref{data_proc}, we provide a statistical characterization of target and input variables, discussing both preprocessing and feature selection. \Cref{models} describes models, including the training process and hyperparameter tuning. In \cref{perf_limit}, we derive the performance limit, which is used as the ultimate benchmark in \cref{results}, where the results are presented and discussed. Finally, \cref{conclusion} is devoted to some concluding remarks.

\mbox{}

\nomenclature{$\mathrm{RGD}(t)$}{Residential Gas Demand at date $t$}
\nomenclature{$\mathrm{\hat{RGD}}(t)$}{one-day-ahead forecast of Residential Gas Demand at date $t$}
\nomenclature{$T(t)$}{temperature in degrees Celsius at date $t$}
\nomenclature{$\hat{T}(t)$}{one day-ahead forecasted temperature in degrees Celsius at date $t$}
\nomenclature{$\mathbf{y}$}{vector of RGD}
\nomenclature{$\mathbf{X}$}{matrix of input features}
\nomenclature{MSCM}{Million of Standard Cubic Meter}
\nomenclature{$\boldsymbol{\beta}$}{vector of model parameters}
\nomenclature{$p(x)$}{probability density function}
\nomenclature{$p(x \mid y)$}{conditional probability density of $x$ given $y$}
\nomenclature{$\kappa(x, y)$}{kernel function}
\nomenclature{$\mathcal{N}(\boldsymbol{\mu},\boldsymbol{\Sigma})$}{multivariate normal distribution with mean $\boldsymbol{\mu}$ and covariance $\boldsymbol{\Sigma}$}
\nomenclature{$\mathcal{D} \otimes \mathcal{W}$}{tensor product of two sets $\mathcal{D}$ and $\mathcal{W}$}
\nomenclature{$\EX[\cdot]$}{expected value}
\nomenclature{$\var{\left[\cdot\right]}$}{variance}
\nomenclature{$\var{\left[\cdot \mid \cdot \right]}$}{conditional variance} 
\nomenclature{$(\textbf{x}_*,y_*)$}{any novel input-output pair}

\printnomenclature
\section{Data} \label{Problem Statement}
In Italy, natural gas is the most common fuel for both power plants and domestic heating. Moreover, several industrial facilities burn gas for either heating or powering productive processes. According to SNAM Rete Gas \cite{snam2017report}, the Italian Transmission System Operator (TSO), in 2017 about 70.59 billions of cubic meters of natural gas were consumed, with an increase of 5.6\% over the previous year. Overall, the increase in demand between 2015 and 2017 was 11\%. Out of the total gas demand in 2017, 35.9\% was due to thermoelectric power plants, 22.4\% to industrial facilities, and 41.7\% to residential users.

Residential Gas Demand (RGD) represents the main part of the overall Italian gas consumption, accounting for household usage for cooking, water heating, and, most importantly, environment heating.

The available dataset covers 11 years, from 2007 to 2017, and is made of 3 fields: date ($t$), forecasted average temperature ($\hat{T}$), and residential gas demand (RGD).
Forecasted temperature is relative to the Northern regions of Italy.
In the preliminary analysis,  a weighted average of the temperatures in different zones of Italy was also considered, but it was dropped because of the weaker correlation with RGD. This is explained by the role of domestic heating in Northern Italy, where winters are colder than in other regions.

\begin{figure}[H]
	\centering
	\includegraphics[width=1\textwidth]{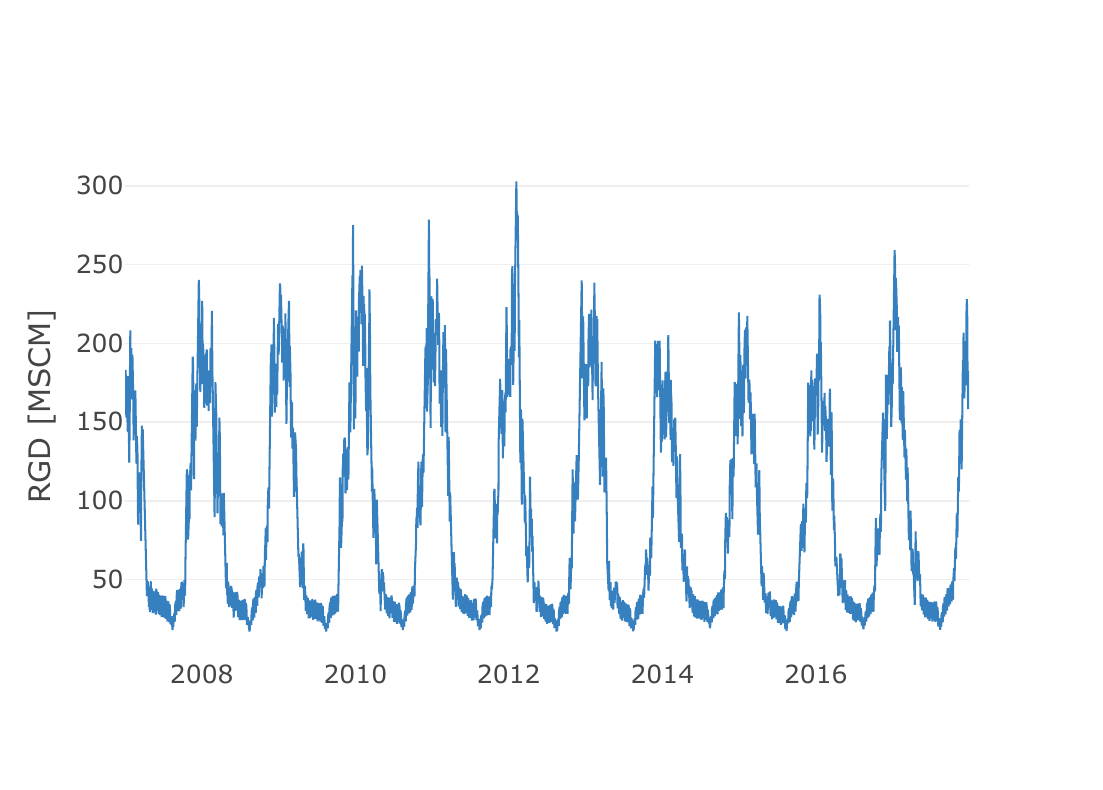}
	\caption{Italian Residential Gas Demand (RGD): years 2007-2017.}
	\label{rgd_series}
\end{figure}

The profile of RGD from 2007 to 2017 is displayed in \cref{rgd_series}.

\section{Exploratory analysis and feature selection}   
\label{data_proc}
\subsection{Residential Gas Demand}
\begin{figure}[H]
	\centering
	\includegraphics[width=1\textwidth]
	{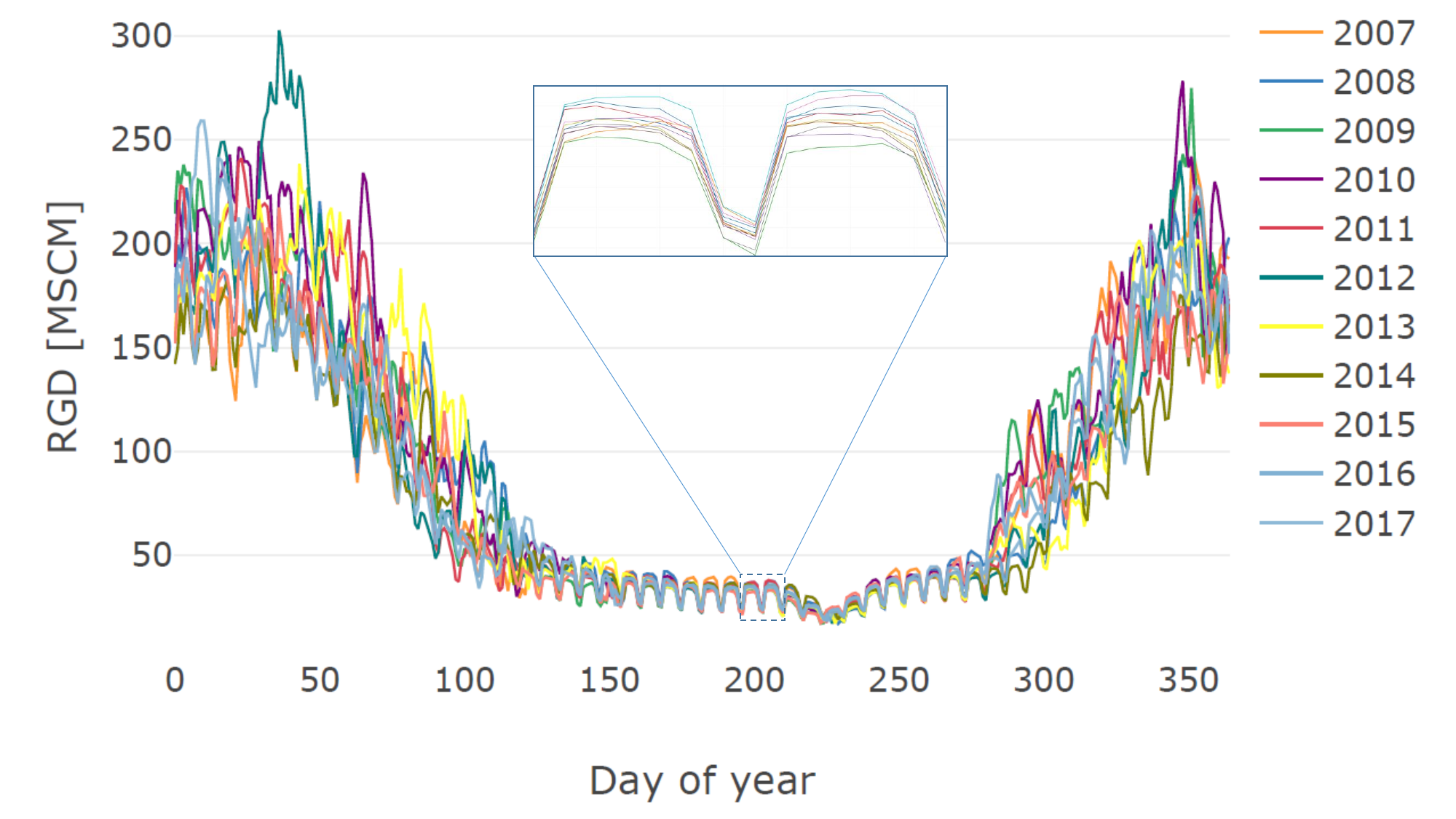}
	\caption{Italian Residential Gas Demand (RGD): years 2007-2017. The time series are shifted to align weekdays: weekly periodicity is particularly visible in summer. The yearly seasonal variation is mostly explained by heating requirements. In the inset, two weeks of July's demand are zoomed.}
	\label{res_superimp}
\end{figure}

RGD magnitude greatly oscillates with the season: during the cold months, from October to March, it represents about 56\% of the overall Italian demand, while it drops to about 28\% during the warm months, from April to September.
In fact, when the temperature climbs above 17-18$^\circ$C, domestic heating is typically switched off. Thus, during the cold period lower temperatures cause a larger RGD, while, during summer, weather influence is negligible and a weekly pattern becomes evident, with lower RGD during weekends compared to working days. Due to the lack of dependence on weather conditions, the profile of summer RGD is remarkably repeatable from year to year. All these features are visible in \cref{res_superimp}, which displays eleven years of Italian RGD, overlapped with a proper shift to align weekdays. 

\begin{figure}[H]
	\centering
	\includegraphics[width=.95\textwidth]{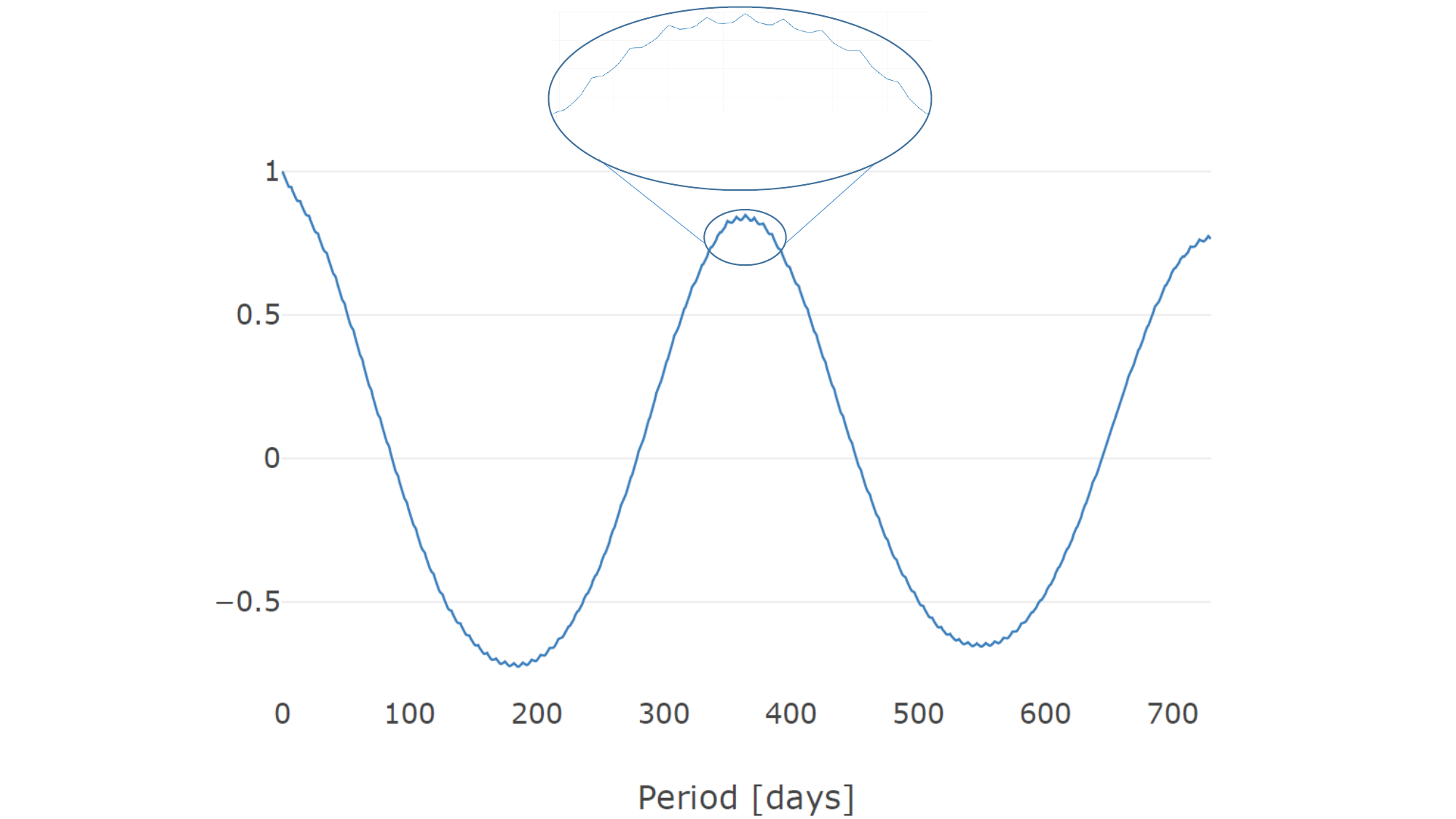}
	\caption{RGD autocorrelation function estimated on 2007-2017 data. The 365-day yearly periodicity is evident. In the inset, weekly waves witness the presence of a 7-day periodicity of smaller amplitude.}
	\label{autocorr}
\end{figure}

As expected, the autocorrelation function, estimated on the whole dataset, exhibits a clear yearly seasonality and a much smaller weekly periodicity, see \cref{autocorr}.

\begin{figure}[H]
	\centering
	\subfloat{\includegraphics[width=.48\textwidth]{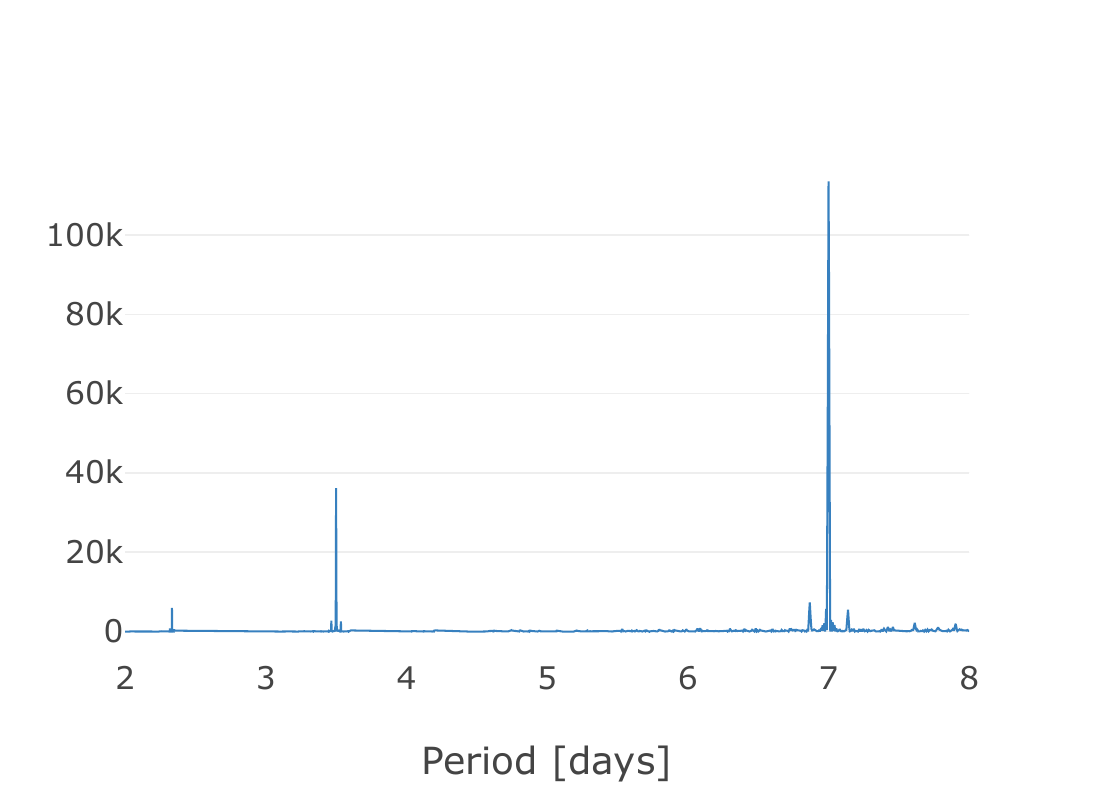}}
	\subfloat{\includegraphics[width=.48\textwidth]{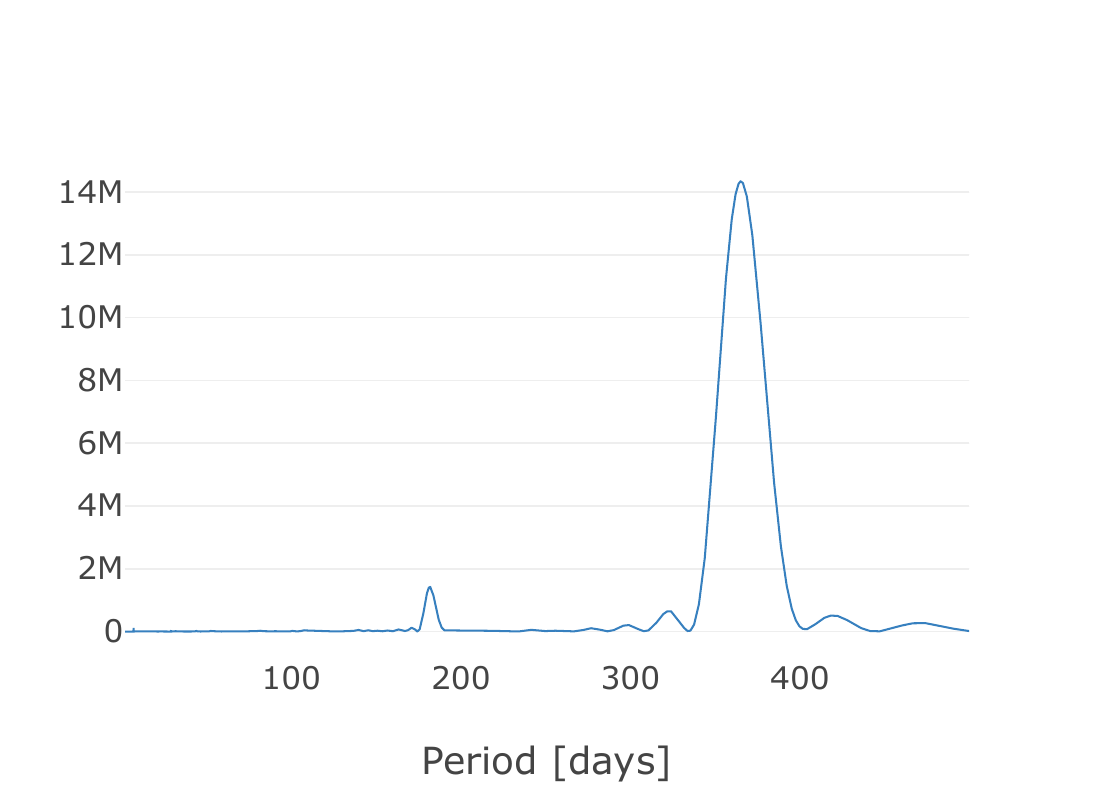}}
	\caption{RGD periodogram. Left panel: periods up to to 8 days; right panel: periods up to 500 days. The yearly periodicity is highlighted by peaks at 365.25 days, while the weekly one by the smaller spike at a period of 7 days. Other spikes correspond to harmonics located at multiples of the main harmonic.}
	\label{spectr_dens}
\end{figure}

Most of the spectral density, see \cref{spectr_dens}, is concentrated at period 365.25 days. A smaller yet relevant peak can be found at a period of 7 days, accounting for the weekly periodicity. In both cases, smaller peaks at lower periods are ascribable to harmonics located at multiples of the main harmonic.

\begin{figure}[H]
	\centering
	\subfloat[$\mathrm{RGD}(t)$ vs $\mathrm{RGD}(t-1)$ \label{scatter t vs t-1}]
	{{\includegraphics[width=0.45\textwidth]{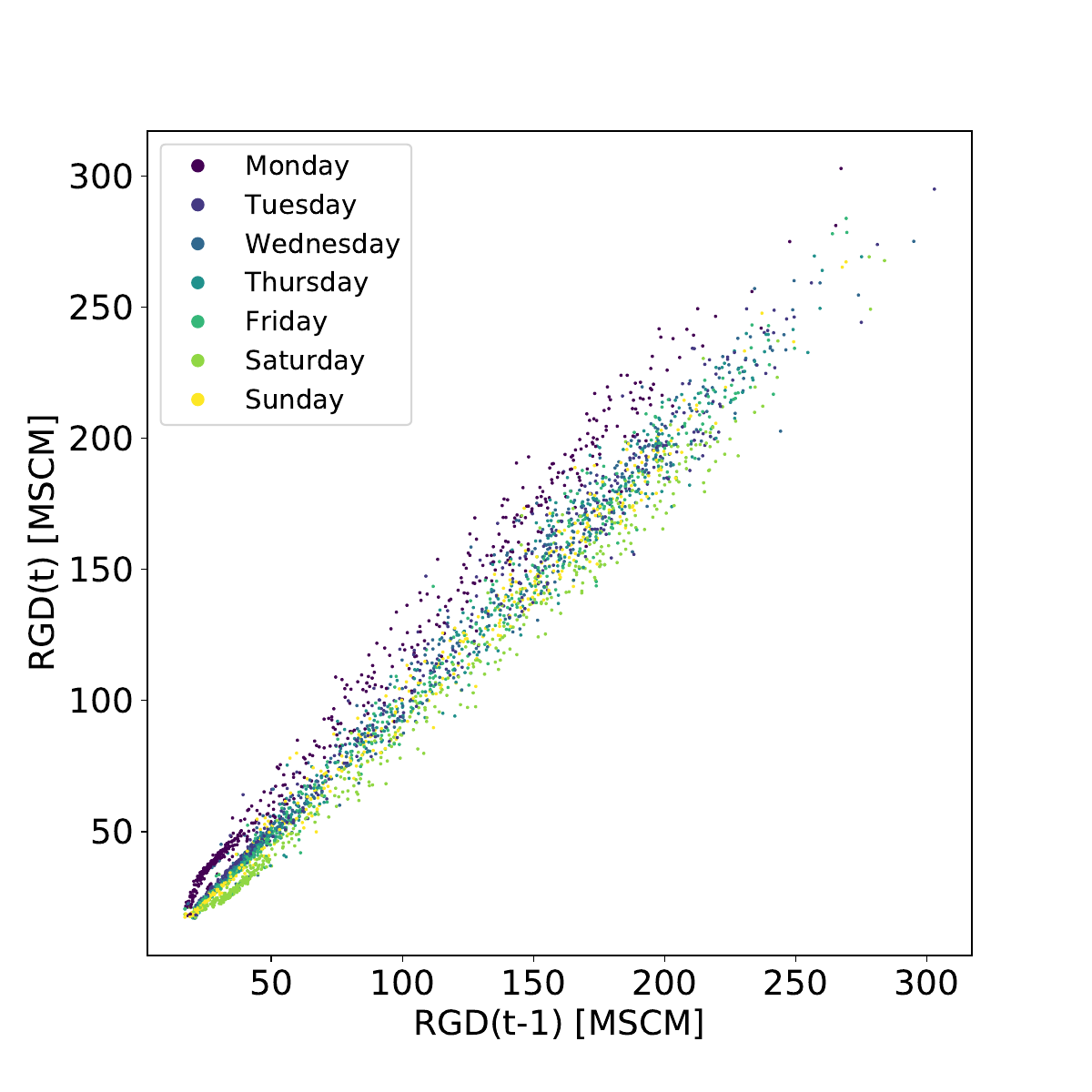} }}%
	\subfloat[$\mathrm{RGD}(t)$ vs $\mathrm{RGD}(t-7)$ \label{scatter t vs t-7}]
	{{\includegraphics[width=0.45\textwidth]{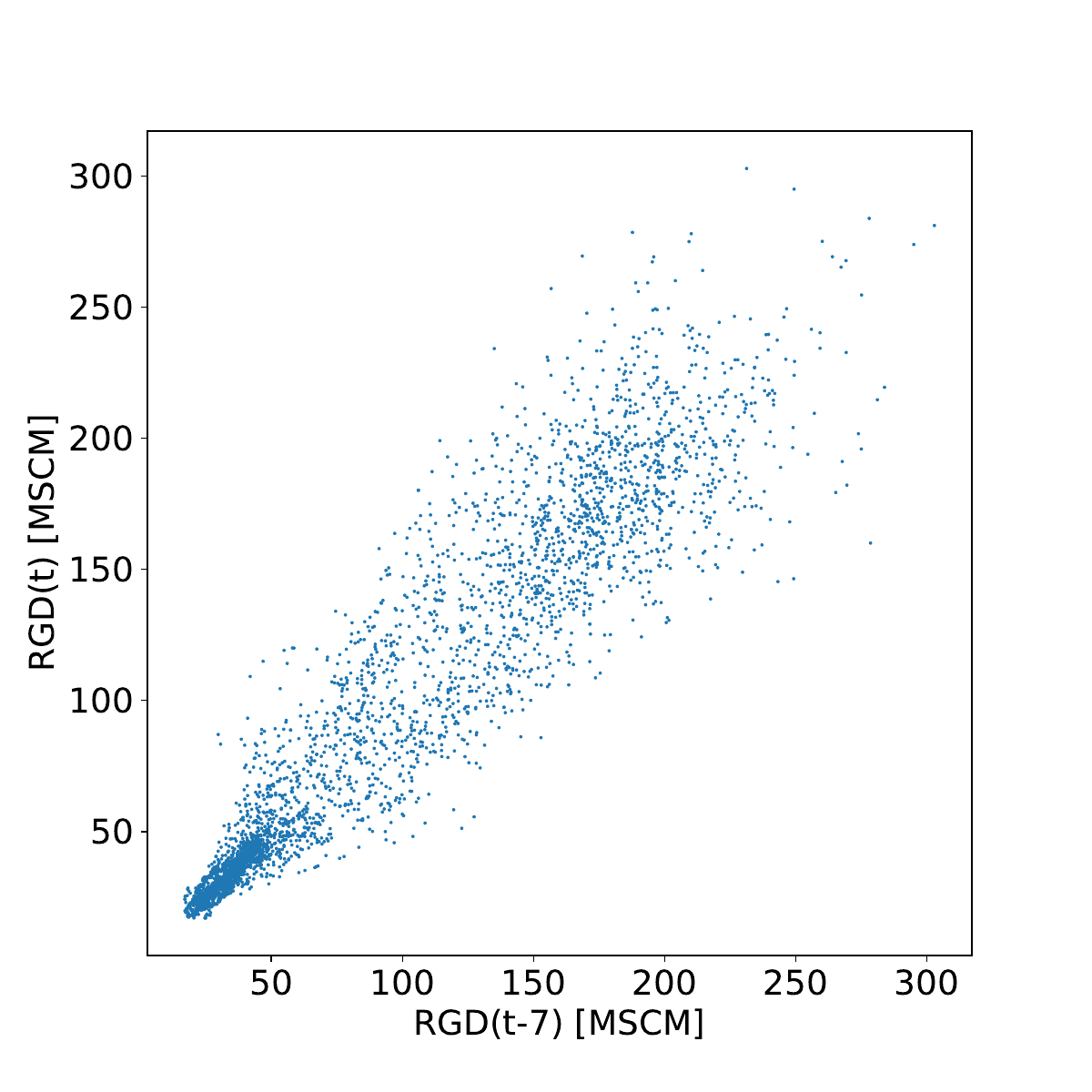} }}%
	\vskip\baselineskip
	\subfloat[$\mathrm{RGD}(t)$ vs $\mathrm{RGD}(\mathrm{sim}(t))$ \label{scatter t vs similar}]
	{{\includegraphics[width=0.45\textwidth]{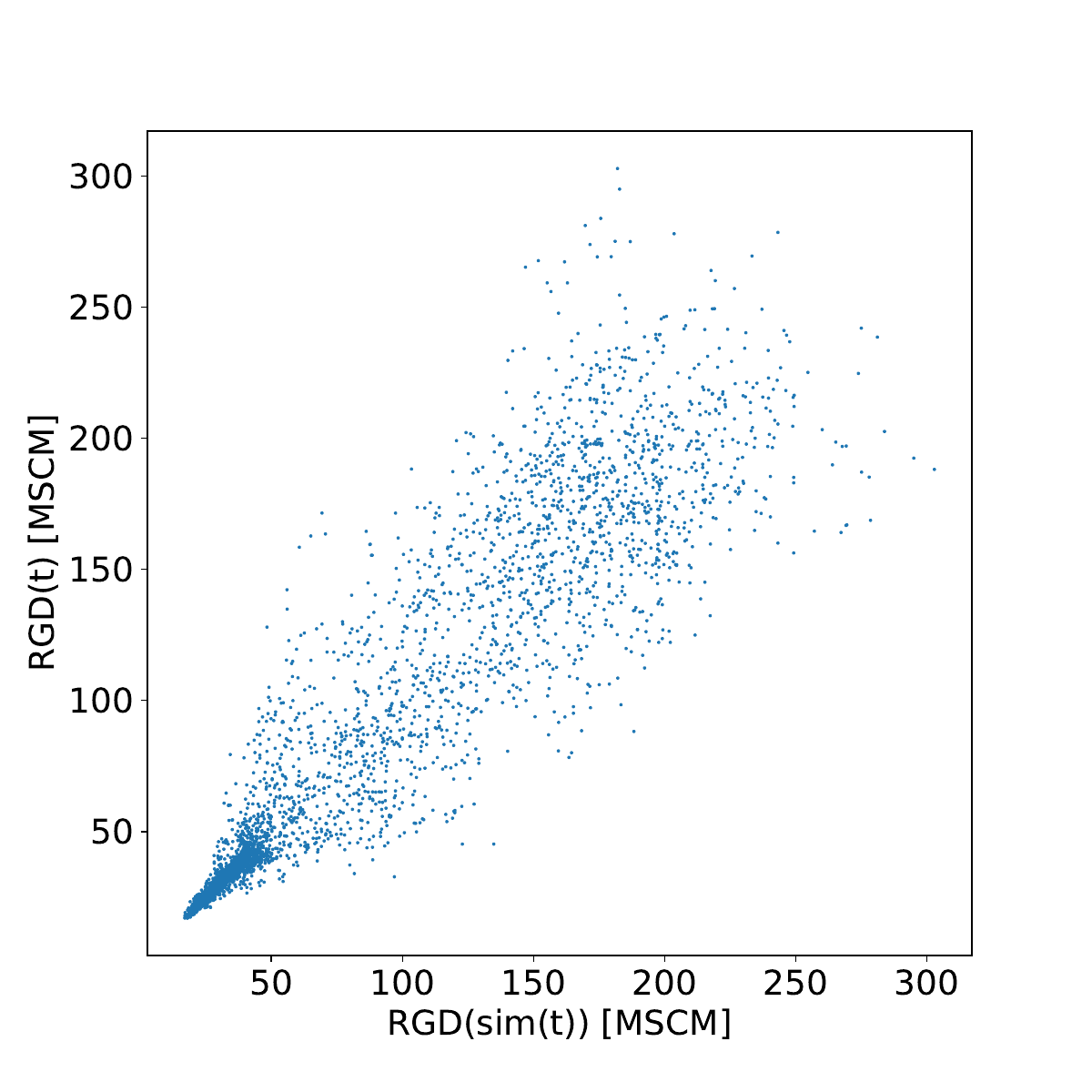} }}%
	\subfloat[$\mathrm{RGD}(t)-\mathrm{RGD}(\mathrm{sim}(t))$ vs $\mathrm{RGD}(t-1)-\mathrm{RGD}(\mathrm{sim}(t-1))$ \label{scatter diff}]
	{{\includegraphics[width=0.45\textwidth]{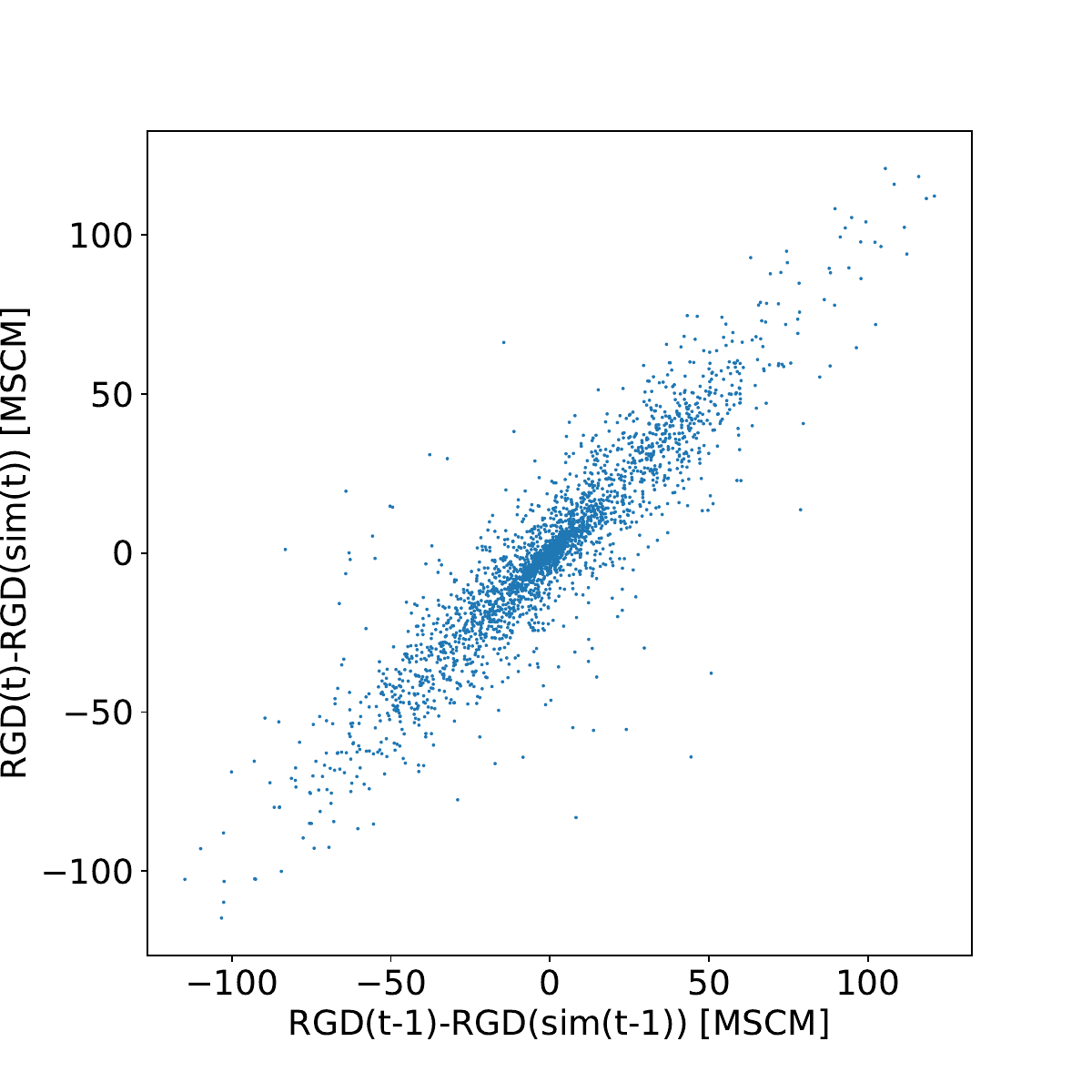}}}
	\caption{Scatter plots involving RGD and potential features to be used for its prediction.}
	\label{scatters}
\end{figure}

The autocorrelation of lag 1 can be assessed through the scatter plot in \cref{scatter t vs t-1}, where RGD at time $t$ is plotted against RGD at time $t-1$. The correlation coefficient computed on the entire dataset is 0.988, and it increases to 0.995 if the pairs Saturday-Friday and Monday-Sunday are discarded. This reflects a different behavior between working days and weekends, visually confirmed in the plot, where Monday's RGD stays in the upper part of the cloud whereas Saturday's RGD lies in the lower part.

As for the lag-7 autocorrelation, in \cref{scatter t vs t-7} the scatter plot of RGD at times $t$ and $t-7$ is displayed. The scatter plot in \cref{scatter t vs t-7} is narrower when the demand is low, that is during warm months, while it gets more dispersed in winter, when the demand is high. This is possibly due to the variability of weather from one week to the next one.

In order to characterize the yearly seasonality, it is convenient to introduce the notion of similar day. The following definitions hold:
\begin{itemize}
\item $\mathrm{year}(t)$ is the year to which day $t$ belongs;
\item $\mathrm{weekday}(t)$ is the weekday of day $t$, e.g. Monday, Tuesday, etc;
\item $\mathrm{yearday}(t)$ is the day number within $\mathrm{year}(t)$ starting from January 1, whose \emph{yearday} is equal to $1$.
\end{itemize}
\begin{definition}[Similar Day]
If $t$ is not a holiday, its similar day $\tau^*=\mathrm{sim}(t)$ is
$$
\tau^* = \arg \min_{\tau} |\mathrm{yearday}(\tau)-\mathrm{yearday}(t)|
$$
subject to
\begin{itemize}
\item $\mathrm{year}(\tau)=\mathrm{year}(t)-1$;
\item $\mathrm{weekday}(\tau)=\mathrm{weekday}(t)$;
\item $\tau$ is not a holiday.
\end{itemize}
If $t$ is a holiday, its similar day $\tau^*=\mathrm{sim}(t)$ is the same holiday in the previous year.
\end{definition}

According to the Italian calendar, holidays are 1 January, 6 January, 25 April, 1 May, 2 June, 15 August, 1 November, 8, 25 and 26 December, Easter and Easter Monday.

The relationship between RGD and RGD in the similar day is shown in \cref{scatter t vs similar}: again, the correlation is higher when the demand is lower, due to the smaller influence of temperature.

It can also be of some interest to take into account the similar day of $t-1$. The scatter plot in \cref{scatter diff} shows that the difference $\mathrm{RGD}(t-1) - \mathrm{RGD}(\mathrm{sim}(t-1))$ is a good proxy to the difference $\mathrm{RGD}(t) - \mathrm{RGD}(\mathrm{sim}(t))$. 

Due to these considerations, $\mathrm{RGD}(t-1)$, $\mathrm{RGD}(t-7)$, $\mathrm{RGD}(\mathrm{sim}(t))$, and $\mathrm{RGD}(\mathrm{sim}(t-1))$ were selected as inputs to forecast RGD at time $t$.

\subsection{Temperature}
\begin{figure}[H]
	\centering
	\includegraphics[width=10cm]{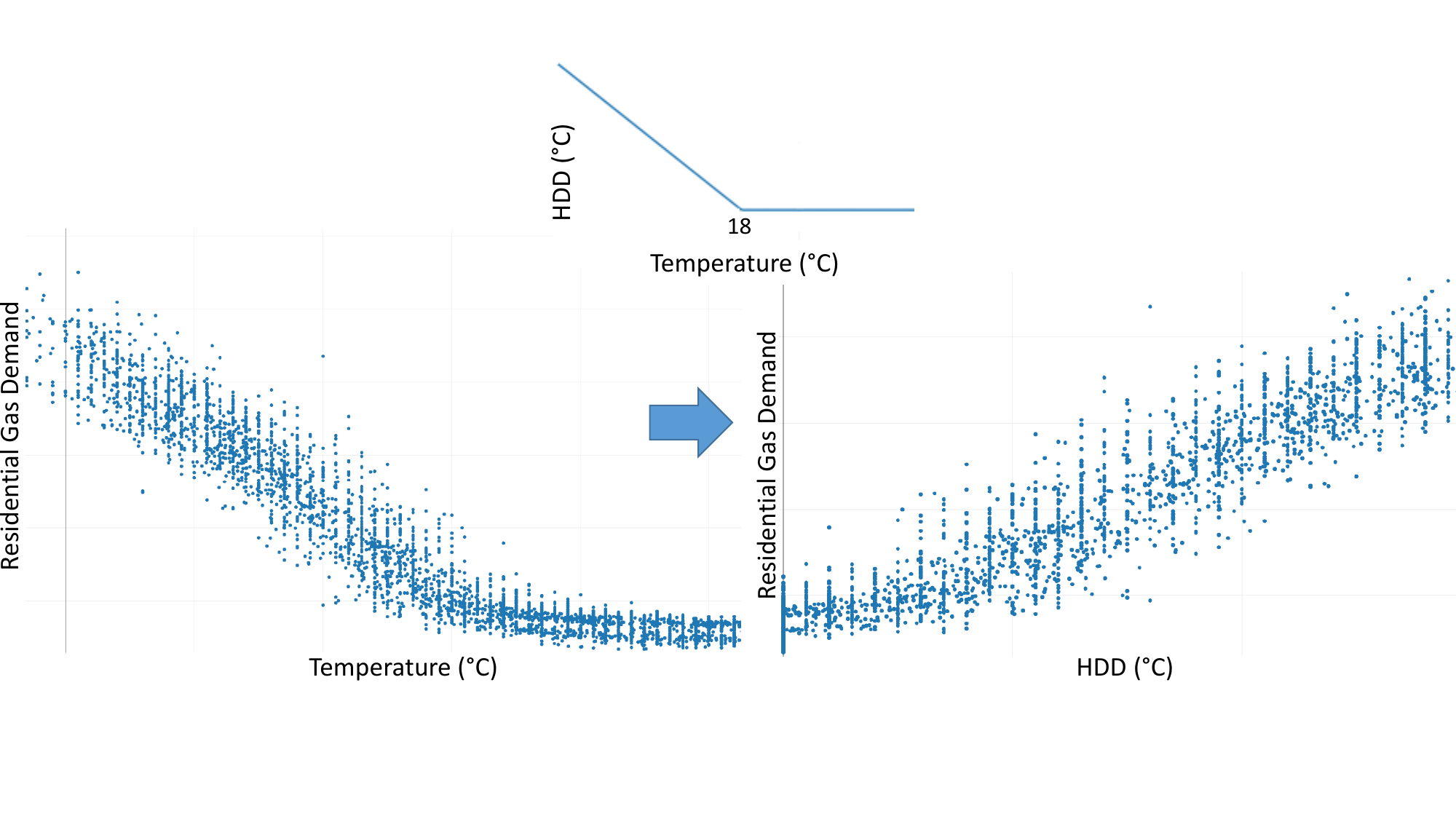}
	\caption{Left panel: scatter plot of daily RGD vs average daily temperature. Right panel: scatter plot of daily RGD vs HDD. Inset: HDD as a function of the temperature.}
	\label{gas_vs_temp}
\end{figure}

The RGD time series shows a strong relationship with temperature, especially during the winter season, when the temperature falls below $18^\circ$C, and household heating becomes relevant. As shown in the left panel of \cref{gas_vs_temp}, with good approximation the relationship is piecewise linear: a line with a negative slope below $18^\circ$C, followed by a constant line above $18^\circ$C. In order to transform the piecewise linear dependence into a linear one, it is useful to refer to the so-called Heating Degree Days (HDD):
\begin{definition} [Heating Degree Days (HDD)]
\begin{equation} \label{HDD}
	\mathrm{HDD}(T)=\max (18^\circ-T,0)
\end{equation}
\end{definition}

\begin{figure}[H]
	\centering
	\includegraphics[width=0.95\textwidth]{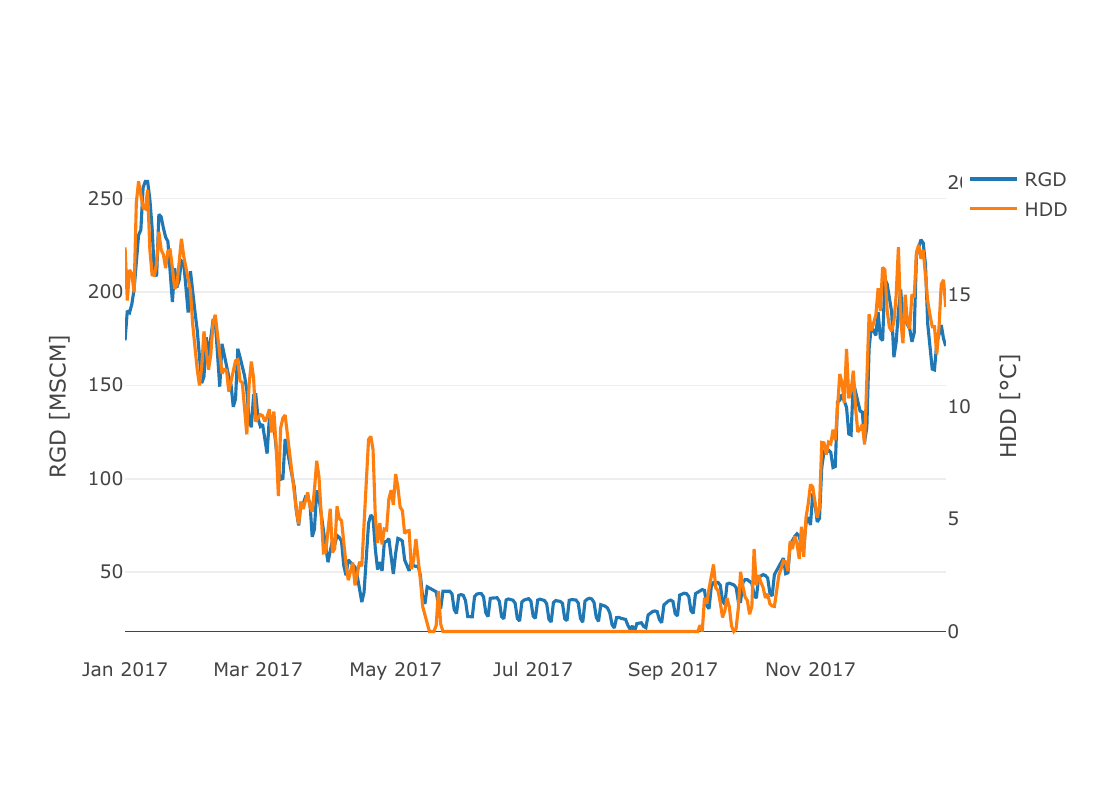}
	\caption{Time series of RGD and HDD in 2017. The instantaneous correlation between the two series is apparent.}
	\label{rgd_vd_hdd}
\end{figure}

In the right panel of \cref{gas_vs_temp}, the scatter plot of RGD vs. HDD highlights an approximately linear relationship, with a positive correlation of $0.97$. The correlation between HDD and RGD is even more evident when we look at the time series of RGD and HDD during 2017, see \cref{rgd_vd_hdd}.

As shown in \cref{gas_vs_temp}, RGD is better correlated to HDD than to plain temperatures. Thus, $\mathrm{HDD}(\hat T(t))$ and the lagged values $\mathrm{HDD}(\hat{T}(t-1)), \mathrm{HDD}(\hat{T}(t-7)), \mathrm{HDD}(\hat{T}(\mathrm{sim}(t)))$ were selected as features, where $\hat T(t)$ denotes the one-day-ahead forecast of $T(t)$. Moreover, the forecasted plain temperature $\hat T(t)$ and its lagged values $\hat{T}(t-1)$, $\hat{T}(t-7)$ and $\hat{T}(\mathrm{sim}(t))$ were also included.

\subsection{Calendar features}
It has already been observed that weekdays and holidays have a great influence on RGD. To capture this phenomenon, the following categorical calendar features were taken into account.

\textit{Weekday}. Given the weekly periodicity, the seven days of the week were taken as explanatory features. By resorting to the dummy encoding method, they were transformed into 6 dichotomic time series.

\textit{Holiday}. A binary feature which takes value $1$ in correspondence of holidays.

\textit{Day after holiday}. A binary feature which takes value $1$ the first \textit{working day} after a holiday. A working day is defined as a day different from Saturday and Sunday that is not a holiday.

\textit{Bridge holiday}. A binary feature which takes value $1$ on isolated working days, that is working days where both the day before and the day after are either Saturday, Sunday or a holiday.

The complete set of features is summarized in \cref{features}.
\begin{table}[h]
\centering

\begin{tabular}{@{}lll@{}}
\toprule
Feature                & Reference time & Type        \\ \midrule \addlinespace
RGD                    & t-1            & continuous  \\
RGD                    & t-7            & continuous  \\
RGD                    & $\mathrm{sim}(t)$         & continuous  \\
RGD                    & $\mathrm{sim}(t-1)$       & continuous  \\ 
\addlinespace
Forecasted temperature & t              & continuous  \\
Forecasted temperature & t-1            & continuous  \\
Forecasted temperature & t-7            & continuous  \\
Forecasted temperature & $\mathrm{sim}(t)$         & continuous  \\
Forecasted HDD         & t              & continuous  \\
Forecasted HDD         & t-1            & continuous  \\
Forecasted HDD         & t-7            & continuous  \\
Forecasted HDD         & $\mathrm{sim}(t)$         & continuous  \\ 
\addlinespace
Weekday                & t              & categorical       \\
Holiday                & t              & binary \\
Day after holiday      & t              & binary \\
Bridge holiday         & t              & binary \\   
\bottomrule

\end{tabular}

\caption{List of features.}\label{features}\end{table}

\section{Prediction models}  \label{models}

The classical methods used for time series forecasting are linear Box-Jenkins models such as SARIMA, where the forecast is based only on past values of the time series, and SARIMAX, that accounts also for exogenous variables. A  drawback of classical linear models is the difficulty in handling the discontinuities due to holidays and the possible presence of other nonlinear phenomena. Below, RGD forecasting is formulated as a statistical learning problem. 

Based on the availability of $n$ data pairs $(\textbf{x}_i,y_i)$, $i=1, \ldots, n$, known as the training data, the goal is to design a prediction rule $f(\cdot)$ with the objective of using $f(\textbf{x}_*)$ as prediction of $y_*$, where $(\textbf{x}_*,y_*)$ is any novel input-output pair. In this context, $\textbf{x}_i \in \mathbb{R}^p$, $p<n$, is a vector whose entries are given by the $p$ features associated with the target $y_i$. 

Herein, the $p$ features are the 21 covariates discussed in the previous section and listed in \cref{features}. In the following, with reference to the training data, $\textbf y=\{y_i\} \in \mathbb{R}^n$ will denote the vector of the targets and $\textbf{X}= \{x_{ij}\} \in \mathbb{R}^{n \times p}$ will denote the matrix of the training input data, where $x_{ij}$ is the $j$-th feature of the $i$-th training pair $(\textbf{x}_i,y_i)$.

Below, five approaches are presented:
\begin{itemize}
    \item ridge regression;
    \item Gaussian Process (GP);
    \item k-nearest neighbour (KNN);
    \item artificial neural network (ANN);
    \item torus model.
\end{itemize}

All the models, except the torus one, were implemented in Python, using scikit-learn and keras packages; automated hyperparameters tuning exploited the GridSearchCV function of scikit-learn.
The torus model was implemented in MATLAB, as well as its hyperparameter tuning routine.

\subsection{Ridge regression}

Ridge regression \cite{hastie2009elements} is a technique to identify a linear model in the form:
\begin{equation*}
\hat{f}(\mathbf{x}) = \sum_{j=1}^{p}x_{j}\beta_j = \mathbf{x}^T \boldsymbol{\beta}, \quad \mathbf{x},\boldsymbol{\beta} \in \mathbb{R}^p
\end{equation*}

To prevent overfitting, besides the standard squared sum of the residuals, the loss function includes also the squared norm of the parameter vector $\boldsymbol{\beta}$:

\begin{equation}
	\boldsymbol{\beta}^{\mathrm{ridge}} := \arg \min_{\mathbf{\beta}} \| \mathbf{y-X} \boldsymbol{\beta} \|^2 +\lambda \|\boldsymbol{\beta} \|^2
	\label{ridge_regr}
\end{equation}

where $\lambda$ is the so-called regularization parameter, a \emph{hyperparameter} which controls the flexibility of the learning algorithm. Assuming that $\mathbf{X}$ is full rank, the solution of \eqref{ridge_regr} is

\begin{equation}
	\boldsymbol{\beta}^{\mathrm{ridge}} = (\mathbf{X}^{T}\mathbf{X}+\lambda \mathbf{I})^{-1} \mathbf{X}^{T}\mathbf{y} 
	\label{ridge_solution}
\end{equation}

that highlights the shrinking effect with respect to the standard least squares estimator $\boldsymbol{\beta}^{LS} = (\mathbf{X}^{T}\mathbf{X})^{-1} \mathbf{X}^{T}\mathbf{y}$. 

Since the parameters are obtained in closed form from \eqref{ridge_solution}, the ridge regression model is entirely specified by choice of $\lambda$, that can be calibrated following different approaches \cite{hastie2009elements}. 
A normalized assessment of the amount of regularization associated with a given $\lambda$ is provided by the so-called effective degrees of freedom
$$
\mathrm{df}(\lambda)=\mathrm{tr} \left(\mathbf{X}(\mathbf{X}^{T}\mathbf{X}+\lambda \mathbf{I})^{-1} \mathbf{X}^{T} \right)
$$
In fact, $\mathrm{df}(\lambda)$ ranges from $p$ to $0$ as $\lambda$ goes from 0 to infinity \cite{hastie2009elements}.

\subsection{Gaussian processes}

Let $ \mathbf{\bar y} =\left[\begin{array}{cc}y_* & \mathbf{y}^T\end{array}\right]^T$, $\mathbf{\bar x} =\left[\begin{array}{cccc}\mathbf{x}_*^{T} & \mathbf{x}_1^T & \ldots &\mathbf{x}_n^T \end{array}\right]^T$ and assume that, conditional on $\mathbf{\bar x}$, the vector $\mathbf{\bar y}$ is normally distributed as follows
\begin{align*}
\mathbf{\bar y} | \mathbf{\bar x} &\sim  \mathcal{N}\left( \mathbf{0},\boldsymbol{\Sigma}(\mathbf{\bar x})+ \sigma^2 \mathbf{I_n} \right )\\
\left[\boldsymbol{\Sigma}(\mathbf{\bar x})\right]_{ij} &= \kappa(\bar x_i,\bar x_j)
\end{align*}
where the \emph{kernel} $\kappa(\cdot,\cdot)$ is a suitable function that specifies the correlation between two target variables $\bar y_i$, $\bar y_j$ as a function of the corresponding input vectors $\mathbf{\bar x}_i$, $\mathbf{\bar x}_j$.  The choice of the kernel by the designer should reflect the available prior knowledge on the characteristics of the prediction rule. It is worth noting that the previous assumption is equivalent to assuming that
\begin{equation*}
y_i = f(\mathbf{x_i}) + \epsilon_i, \quad i=1, \ldots, n
\end{equation*}
where $\epsilon_i \sim \mathcal{N}(0, \sigma^2)$ are independent errors and $f(\cdot)$ is the realization of a zero-mean continuous-time Gaussian Process (GP) with autocovariance $\kappa(\mathbf{\bar x_i},\mathbf{\bar x_j})$\cite{williams2006gaussian, murphy2012machine}.
The estimation of a new target value $y_*$ relies on the following well known property of normally distributed random vectors.
\begin{lemma}[Conditional distribution of jointly Gaussian variables]
Let $z_*$ and $\mathbf{z}$ be zero-mean jointly Gaussian random variables distributed as follows:
\begin{equation*}
    \begin{bmatrix}z_*\\\mathbf{z}\end{bmatrix} \sim \mathcal{N} \left(
    \begin{bmatrix}0\\\mathbf{0}\end{bmatrix},
    \begin{bmatrix}
        \boldsymbol{\Sigma}_{z_*z_*} + \sigma^2  & \boldsymbol{\Sigma}_{z_*z}\\
        \boldsymbol{\Sigma}_{zz_*}   & \boldsymbol{\Sigma}_{zz} + \sigma^2 \mathbf{I_n}
    \end{bmatrix}
    \right)
\end{equation*}
Then, the posterior distribution of $z_*$ conditional on $\mathbf{z}$ is:
\begin{equation*}
    z_* | \mathbf{z} \sim \mathcal{N}\left( \Sigma_{z_*z}\left(\mathbf\Sigma_{zz}+\sigma^2\mathbf I_n\right)^{-1}\mathbf{z}, \; \Sigma_{z_*z_*} + \sigma^2 - \mathbf \Sigma_{z_*z}\left(\mathbf \Sigma_{zz}+\sigma^2\mathbf I_n\right)^{-1}\mathbf\Sigma_{zz_*} \right)
\end{equation*}
\end{lemma}

In view of the previous lemma, it is possible to use the posterior expectation as prediction rule:
\begin{align*}
f(\mathbf{x_*})&= \EX \left[y_* | \mathbf{x_*}, \mathbf{y}, \mathbf{x} \right]= \sum_{i=1}^n c_i \kappa(\mathbf{x_*},\mathbf{\bar x_i})\\
 \boldsymbol{c} &= \left( \boldsymbol{\Sigma}(\mathbf{x}) + \sigma^2 \mathbf{I_n} \right)^{-1} \boldsymbol{y}
\end{align*}

The main distinctive feature of GP models is the learning process, which aims directly at obtaining the predictive function rather than inferring its parameters.

A zero-mean GP is completely defined by its covariance function $\kappa(\mathbf{ x_i},\mathbf{x_j})$, also called kernel. When it is a function of the distance $r= \| \mathbf{x_i}-\mathbf{x_j} \|$ between $x_i$ and $x_j$, i.e. $\kappa(\mathbf{x_i}, \mathbf{x_j})=\kappa(r)$, the kernel is said to be stationary and isotropic. Within this class, a popular and flexible choice is the family of Matérn kernels, defined by:

\begin{equation*}
	\kappa_{\text{Matern}}(r)=\frac{2^{1-\nu}}{\Gamma(\nu)}\Bigl(\frac{\sqrt{2\nu}r}{l}\Bigr)^{\nu}K_{\nu}\Bigl(\frac{\sqrt{2\nu}r}{l}\Bigr)
\end{equation*}

where $\nu$ and $l$ are hyperparameters to be tuned and $K_{\nu}$ is a modified Bessel function \cite{abramowitz1948handbook}. The parameter $l$ defines the characteristic length-scale of the process, whereas $\nu$ defines the specific covariance function in the Matérn class. 





Different approaches are possible in order to tune the hyperparameters $\nu, \lambda$, and $\sigma^2$. According to an empirical Bayes, the hyperparameter vector $\boldsymbol{\eta}$ is chosen as the maximizer of the marginal likelihood $p(\mathbf{y}|\mathbf{x},\boldsymbol{\eta})$.

\subsection{K-Nearest neighbors}

K-Nearest neighbours (KNN) relies on the distance between samples in the feature space: given a test sample $\mathbf{x_*}$, the prediction of $y_*$ is computed by averaging $K$ training samples $y_i, i \in \mathcal C$, where $\mathcal C$ denotes the set identified by the $K$ feature vectors $\mathbf{x_i}$ that are closest to $\mathbf{x_*}$, according to some distance measure, e.g., the Euclidean norm (herein adopted).

In order to specify a KNN estimator, one has to choose the distance metric, e.g., Euclidean, Minkowsky, Manhattan, etc, and the type of weighted average, e.g., uniform or inverse distance, and to calibrate one hyperparameter, viz the number $K$ of neighbors. Too small values of $K$ lead to overfitting to the training data, while including too many neighbors reduces the variance at the cost of jeopardizing model flexibility. 

\subsection{MLP Neural Network}
Artificial Neural Networks (ANN) are complex non-linear models, capable of capturing non-linear patterns and relations. A comprehensive explanation of their structure and the most common training algorithms can be found in \cite{Goodfellow2016deep}.

In this study, we focused on the Multi-Layer Perceptron (MLP) or fully connected ANN. 


 The Rectified Linear Unit (ReLu) activation function and the Mean Squared Error (MSE) loss function were adopted. The training was performed using gradient descent, as implemented in the Adaptive Moment Estimation (ADAM) algorithm \cite{kingma2014adam}. The hyperparameters to be tuned include the number of neurons in each layer, the parameters entering the definition of the activation functions, and optimization parameters such as the number of epochs, batch size, and learning rate.

\subsection{Torus model}

The torus model \cite{guerini2015long} is a linear model based on sinusoidal functions, developed initially to predict electrical power load.
Its short-term version was adapted in order to forecast RGD.

Following \cite{guerini2015long}, a logarithmic transformation of the RGD was performed in order to mitigate the effect of its skewness. The long-term model is

\begin{equation*}
\ln	\hat{\mathrm{RGD}}^{\it{long}}(t)=L(t)+F(t)+\sum_{i}H_{i}(t)
\end{equation*}

where the log-forecast is the sum of three elements: the trend or level $L$, the multiperiodic term $F$, which accounts for seasonality, and the effect of holidays $\sum_{i}H_{i}$.

The multiperiodic term $F$ is modeled by a linear combination of sinusoidal functions:

\begin{equation*}
	F(t)=\sum_{i=1}^{(1+2N_d)(1+2N_{w})}\theta_{i}h_{i}(t)
\end{equation*}

where the functions $h_i$ are given by the product of the $j$-th element in $\mathcal{D}$ with the $k$-th element in $\mathcal{W}$, for suitable $j$ and $k$, with  

\begin{equation}
	\mathcal{D}=\{\cos(j\Psi t),j \in [0, N_{d}]\} \cup \{\sin(j\Psi t),j=\in [1, N_{d}] \} \nonumber
\end{equation}
\begin{equation}
	\mathcal{W}=\{\cos(k\Omega t),k\in [0, N_{w}]\} \cup \{\sin(k\Omega t),k \in [1, N_{w}] \} \nonumber 
\end{equation}

The frequencies of the sinusoidal functions are $\Psi =\frac{2\pi}{365.25}$ and $\Omega =\frac{2\pi}{7}$. The number of harmonics, respectively $N_w$ for 7-day and $N_d$ for 365.25-day periodicity, are hyperparameters of the model.

The dependency on temperature, expressed in $\mathrm{HDD}(t)$, and its daily difference $\mathrm{HDD}(t)-\mathrm{HDD}(t-1)$, was accounted for by including these two features in the set of regressors. The terms related to trend and holidays were the same as in \cite{guerini2015long}.

Finally, to get a short-term predictor, the long-term model was corrected with the gas demand of the previous day:

\begin{equation*}
	\hat{\mathrm{RGD}}(t)=\hat{\mathrm{RGD}}^{\it{long}}(t)\frac{\mathrm{RGD}(t-1)}{\hat{\mathrm{RGD}}^{\it{long}}(t-1)}.
\end{equation*}

The number of harmonics $N_w$ and $N_d$ were tuned by minimizing the AIC index.

\section{Effects of temperature forecast errors} \label{perf_limit}
As shown in \cref{data_proc}, temperature is the most important exogenous variable. Obviously, the actual temperature cannot practically be used when forecasting future RGD: only a forecast is available, affected by a small yet non-negligible error, which inevitably impacts also the performance of  gas demand forecast. The scope of this section is to assess the influence of the temperature error on the precision of the RGD forecast. For this purpose, we resort to an idealized error propagation model that, despite its simplicity, provides an accurate description, as confirmed by the subsequent experimental validation.

Let RGD be a deterministic function $g$ of the true temperature $T$ and some other factors $\mathbf{x} = \left(x_1, x_2, ...\right)$: $\mathrm{RGD} = g\left(T, \mathbf{x}\right)$. In view of the analysis and the charts presented in \cref{data_proc}, a first-level approximation of the relationship between RGD and $T$ is a linear function of HDD, while the dependence on the other factors can be represented as an additive term $\bar{g}(\mathbf{x})$:
\begin{equation*}
	\mathrm{RGD} = g\left(T, \mathbf{x}\right) =\bar{g}\left(\mathbf{x}\right) + \alpha \mathrm{HDD}(T)
\end{equation*}
where $\alpha$ is the sensitivity of the gas demand to HDD. The formula is of general validity and applies to both regional and national gas markets. Indeed, $\alpha$ depends on the size of the considered market and can be estimated from historical data, e.g. those displayed in \cref{gas_vs_temp}.

Consider now the ideal case when $\alpha$ and also the function $\bar{g}$ are perfectly known, yet, only a forecast
$$
\hat{T} = T + \epsilon
$$
of the correct temperature $T$ is available, where $\epsilon$ is a zero-mean error with variance $\sigma^2_\epsilon$.
The optimal forecast $\hat{\mathrm{RGD}}$, given $\hat{T}$, is therefore:
\begin{equation*}
	\hat{\mathrm{RGD}} =  \bar{g}\left(\mathbf{x}\right) + \alpha \mathrm{HDD}(\hat{T}) 
\end{equation*}
In order to obtain the mean squared error of $\hat{\mathrm{RGD}}$, we first compute the conditional variance of $\hat{\mathrm{RGD}}$:
\begin{align*}
	\var{\left[\hat{\mathrm{RGD}} \mid T \geq 18^\circ \right]} & 
	= \var\left[{\bar{g}\left(\mathbf{x}\right) + \alpha \cdot 0}\right]
	= 0 \\
	\var{\left[\hat{\mathrm{RGD}} \mid T < 18^\circ \right]} &
	= \var\left[{\bar{g}\left(\mathbf{x}\right) + \alpha\left(18^\circ - \hat{T} \right)}\right]
	= \alpha^2\var{\left[\epsilon\right]}
	= \alpha^2\sigma^2_\epsilon
\end{align*}
Since $\EX[\epsilon]=0$, it follows that $\EX[\hat{\mathrm{RGD}}] = \mathrm{RGD}$. Thus:
\begin{align} \label{perflim}
    \EX\left[\left(\hat{\mathrm{RGD}} - \mathrm{RGD}\right)^2\right] &=
	\EX\left[\left(\hat{\mathrm{RGD}} - \mathrm{RGD}\right)^2 \mid T \geq 18^\circ \right]P\left(T\geq 18^\circ \right) + \nonumber \\ 
	&+\EX\left[\left(\hat{\mathrm{RGD}} - \mathrm{RGD}\right)^2 \mid T < 18^\circ \right]P\left(T < 18^\circ \right)= \nonumber\\ 
	&=0 + \var{\left[\hat{\mathrm{RGD}} \mid T < 18^\circ \right]} =\nonumber \\ 
	&=P\left(T < 18^\circ\right)\alpha^2\sigma^2_\epsilon
\end{align}
This last equation provides an estimate of the mean squared error due to the temperature forecasting error. Since it has been derived under an ideal setting, i.e. $\alpha$ and $\bar{g}(\cdot)$ perfectly known, it provides a lower limit to the precision that can be achieved by the best possible forecaster.

\begin{figure}[H]
	\centering
	\includegraphics[width=0.8\textwidth]{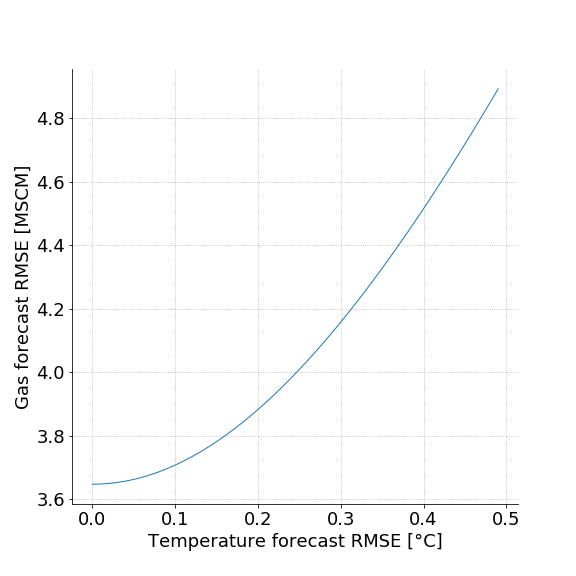}
	\caption{The relation between the Gas forecast RMSE and the Temperature forecast RMSE.}
	\label{temperature_effect_on_gas_forecast}
\end{figure}

The arguments entering the bound are easily obtainable as follows:
\begin{itemize}
    \item Estimate $P\left(T < 18^\circ\right)$ by computing the ratio between the number of samples such that $T < 18^\circ$ and the total number of available data.
    \item Compute $\alpha$ through a least square fit of $\mathrm{RGD}$ vs $T$.
    \item Estimate $\sigma^2_\epsilon$ as the sample variance of $\hat T - T$. 
\end{itemize}
Considering the Italian RGD data, in the 3-year period 2015-2017, $P\left(T < 18^\circ\right)$ ranges from 54\% to 67\%, while $\sigma^2_\epsilon$ ranges from 0.05 to 0.09, and $\alpha$ from 9.85 to 10.96. Considering altogether the years 2015-2017, we have $P\left(T < 18^\circ\right)=63\%$, $\sigma^2_\epsilon = 0.063$, $\alpha=10.56$, corresponding to a best achievable Root Mean Squared Error
$$
\mathrm{RMSE}=\sqrt{	\EX\left[\left(\hat{\mathrm{RGD}} - \mathrm{RGD}\right)^2\right] }=10.56 \times \sqrt{0.63\times 0.063} =2.22\enspace \mathrm{MSCM}
$$
Finally, we consider the more realistic case in which the forecasting mean square error is different from zero even in absence of temperature errors, that is 
$$
\var{\left[\bar{g}\left(\mathbf{x}\right)\right]}= \sigma_0^2 > 0
$$
Then, under a statistical independence assumption, it is possible to obtain the forecasting RMSE as a function of $\sigma^2_\epsilon$:
\begin{equation} \label{tempeffect}
    \mathrm{RMSE}(\sigma^2_\epsilon)=\sqrt{\var{\left[\hat{\mathrm{RGD}}\right]}} =\sqrt{\sigma_0^2+P\left(T<18^\circ\right)\alpha^2\sigma^2_\epsilon}
\end{equation}
In \cref{temperature_effect_on_gas_forecast} this relationship is displayed assuming  $P\left(T < 18^\circ\right)=63\%$, $\sigma^2_\epsilon = 0.063$, $\sigma_0^2=13.31$ (this last value is the test MSE achieved by the ANN forecaster trained with true temperature data instead of the forecasted ones, see \cref{results}). Notably, the sensitivity of the gas forecasting error tends to increase as the temperature forecast error grows. In particular, if we define the threshold
$$
\bar \sigma^2_\epsilon = \frac{\sigma_0^2}{P\left(T<18^\circ\right)\alpha^2}
$$
the influence of temperature errors is negligible as far as 
$\sigma^2_\epsilon \ll \bar \sigma^2_\epsilon$, while the temperature errors have a linear influence on the gas RMSE for $\sigma^2_\epsilon \gg \bar \sigma^2_\epsilon$.

\section{Results} \label{results}
\subsection{Data and performance indicators}
As mentioned in \cref{Problem Statement}, available data range from 2007 to 2017. 
Three one-year-long test sets were defined, associated with the year 2015, 2016, and 2017. The corresponding three training sets spanned from 2007 to the day before the start of the test set: 2007-2014, 2007-2015, 2007-2016. In the following, each training set is identified by the year of the corresponding test set, e.g., we will write "training set 2016" to indicate the second training set, spanning from 2007 to 2015.

On each test set, the performance of the five models was measured using the Mean Absolute Error (MAE).
\begin{equation*}
	\mathrm{MAE} = \frac{1}{N}\sum\limits_{j=1}^{N}\left|\mathrm{RGD}_j-\hat{\mathrm{RGD}}_j\right|
\end{equation*}
MAE is preferred over MAPE due to the highly non-stationary behavior of RGD series. Using MAPE would attribute undue importance to errors during the summer period when RGD is small, see \cref{res_superimp}.  Moreover, MAE is proportional to the monetary loss sustained by energy companies because of errors in nomination due to inaccurate forecasts.

Nevertheless, in order to allow a comparison with forecasting  performances achieved in the UK market, we will also refer to the MAPE:
$$
\mathrm{MAPE} = \frac{100}{N}\sum\limits_{j=1}^{N}\frac{\left|\mathrm{RGD}_j-\hat{\mathrm{RGD}}_j \right|}{\mathrm{RGD}_j}
$$
Indeed, the comparison between two different markets calls for the use of a relative metric. To avoid the confounding effect of small absolute errors that are amplified by MAPE during the Italian Summer, the comparison between Italy and UK was limited to the cold months, when gas demand is relatively high, see Section \ref{Prediction_results}.

Finally, as the performance limit derived in \cref{perf_limit} poses a lower bound to the mean squared error, the Root Mean Square Error (RMSE) was also used as a comparison metric.

\subsection{Hyperparameters}

All five models include hyperparameters that were tuned by cross-validation.

For ridge regression, the regularization parameter $\lambda$ was tuned by 5-fold cross-validation on an interval ranging from $10^{-4}$ to $10^2$ in logarithmic steps. In the training set 2015, line search selected $\lambda=0.236$, corresponding to $\mathrm{df}(0.236)=20.94$, while in the other two sets, 2016 and 2017, cross-validation selected the minimum $\lambda=10^{-4}$ with effective degrees of freedom $\mathrm{df}(10^{-4})=20.99$ practically equal to the number of parameters. This means that regularization plays a very marginal role.

For KNN, we optimized the number of neighbors in the interval $[1,30]$ as well as the weighting strategy, choosing between uniform and inverse of the distance. We obtained seven neighbors for training set 2015 and 6 for the two remaining ones. In all the three cases, the "inverse of distance" weights were selected.

As for the Gaussian Process, the maximization of the marginal likelihood yielded $\nu=1.5$, $l=10$, and $\sigma^2=10$, with minimal variations among all training sets.

For the ANN models, a trial and error procedure led to an architecture with an input layer of 24 neurons, two hidden layers of 12 and 4 neurons, and an output layer of a single neuron. More complex structures led to overfitting and loss of predictive performance. By 5-fold cross-validation, we obtained a learning rate of 0.001 and a batch size of 32. We selected 1000 epochs for training by observing the evolution of the loss function on train and validation sets. 

For what concerns the Torus model, the minimization of AIC led to the choice of $N_w=3$ and $N_d=1$ for all the training sets.

\subsection{Prediction results}
\label{Prediction_results}

A first assessment of the performances of the adopted methods was carried out in terms of RMSE. In order to validate the formula that models  the propagation of temperature errors (\cref{perf_limit}), two sessions were performed. In the first one, the models were trained and tested using historical records of true temperatures, assuming that the one-day-ahead exact temperature is available as a feature. Then, we used \cref{tempeffect} in order to predict how much the forecasting RMSE would increase in the more realistic scenario in which one-day-ahead temperature forecasts are employed in place of the true temperatures. In the second session, the models were trained and tested using historical records of forecasted temperatures. In this way, it was possible to validate the error propagation model against the real errors.

The results of the first session are summarized in \cref{models_with_temp_cons}. It can be seen that the smallest RMSE is obtained by GP and ANN, the latter being marginally better. 

\begin{table}[!htbp]
	\centering
		\begin{tabular}{@{}lcccc@{}} 
		\toprule
			Year            & 2015 & 2016 & 2017 & 2015-2017\\
		\midrule \addlinespace
			Ridge            & 4.24      & 4.11      & 4.12     & 4.16  \\
			GP               & 3.81      & 3.68      & 3.64     & 3.71  \\
			KNN              & 7.29      & 8.49      & 8.38     & 8.07 \\
			Torus            & 4.21      & 4.23      & 3.70     & 4.05  \\
			ANN              & 3.89      & 3.60      & 3.44     & 3.65  \\
		\bottomrule
		\end{tabular}
	\caption{Performance on test sets: yearly RMSE (MSCM) of the five forecasters trained and tested assuming that one-day-ahead true temperatures are available.}
	\label{models_with_temp_cons}
\end{table}

Obviously, in a real-world context, only temperature forecasts are available for the day ahead. In order to account for the performance degradation due to the use of forecasted temperatures, \cref{tempeffect} was used to predict the RMSE of the RGD forecast in correspondence of a temperature forecast variance $\sigma^2_\epsilon = 0.063$, coinciding with that of our meteorological data. The results are summarized in \cref{th_forecasts}. In the first line, the theoretical performance limits computed according to \cref{perflim} are reported. These values were added to the RMSE's of \cref{models_with_temp_cons} to obtain predictions of RGD forecasting RMSE in a real-world situation in which one-day-ahead temperature forecasts are used.

\begin{table}[!htbp]
	\centering
		\begin{tabular}{@{}lcccc@{}} 
		\toprule
			Year            & 2015 & 2016 & 2017 & 2015-2017 \\
		\midrule \addlinespace
			Performance limit   & 2.15      & 2.02      & 1.98  & 2.05 \\
			Ridge            & 4.75      & 4.58      & 4.57   & 4.63   \\
			GP               & 4.37      & 4.20      & 4.15  & 4.24    \\
			KNN              & 7.60      & 8.73      & 8.61  & 8.33    \\
			Torus            & 4.73      & 4.69      & 4.20  & 4.55    \\
			ANN              & 4.45      & 4.13      & 3.97  & 4.19    \\
		\bottomrule
		\end{tabular}
	\caption{Predicted performance on test sets when temperature forecasts with $\sigma^2_\epsilon = 0.063$ are used: yearly RMSE (MSCM) of the five forecasters.}
	\label{th_forecasts}
\end{table}

\begin{figure}[H]
	\centering
	\includegraphics[width=0.8\textwidth]{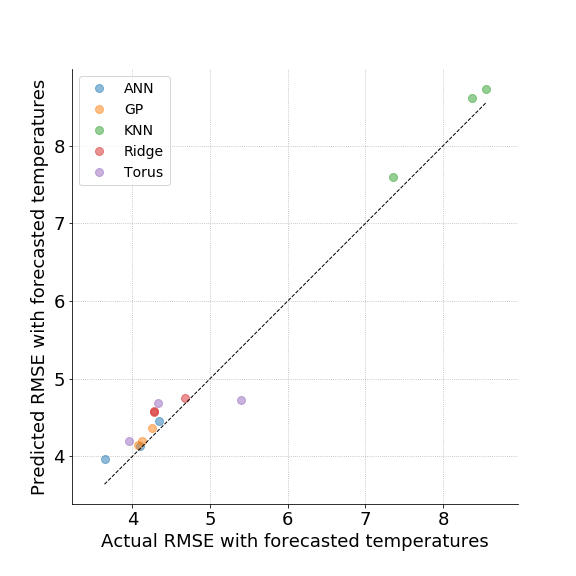}
	\caption{Validation of the model predicting effects on gas forecast of temperature forecast errors. Gas forecast RMSE: theoretical prediction vs actual value.}
	\label{actual_vs_simulated_rmse}
\end{figure}

In the second session, the predictions of \cref{th_forecasts} were validated by comparing them with the RGD forecasting RMSE achieved using temperature forecasts. As it can be seen in \cref{models_vs_perf_limit}, the actual RMSE are in good agreement with their predictions. This can also be visually appreciated in \cref{actual_vs_simulated_rmse}, where theoretical predictions are plotted against the actual RMSE. 




\begin{table}[!htbp]
	\centering
		\begin{tabular}{@{}lcccc@{}} 
		\toprule
			Year            & 2015 & 2016 & 2017 & 2015-2017\\
		\midrule \addlinespace
			Ridge            & 4.68      & 4.28      & 4.28   & 4.42   \\
			GP               & 4.25      & 4.12      & 4.07   & 4.15   \\
			KNN              & 7.35      & 8.55      & 8.37   & 8.11   \\
			Torus            & 5.40      & 4.33      & 3.96   & 4.60   \\
			ANN              & 4.34      & 4.10      & 3.64   & 4.04   \\
		\bottomrule
		\end{tabular}
	\caption{Performance on test sets: yearly RMSE (MSCM) of the five forecasters trained and tested using one-day-ahead forecasted temperatures.}
	\label{models_vs_perf_limit}
\end{table}

A second assessment of the models was made in terms of their MAE. Hereafter, one-day-ahead forecasted temperatures are employed in the features. Results on the test sets are shown in \cref{mae}.
Now, GP is the best performer, achieving an average MAE of 2.53 MSCM over the three test years. ANN, Torus, and Ridge Regression follow in the order. KNN is again the worst model, with an average MAE of 5.05 MSCM.

\begin{table}[!htbp] 
	\begin{center}
		\begin{tabular}{@{}lcccc@{}} 
			\toprule
			Year   &  2015 & 2016 & 2017 & 2015-2017 \\
			\midrule \addlinespace
			Ridge &  3.39      & 3.10      & 3.01      & 3.17    \\
			GP    &  2.60      & 2.48      & 2.51      & 2.53    \\
			KNN   &  4.57      & 5.51      & 5.08      & 5.05    \\
			Torus &  3.18      & 2.66      & 2.55      & 2.80    \\
			ANN   &  2.76      & 2.68      & 2.43      & 2.62    \\
			\bottomrule
		\end{tabular}
	\end{center}
	\caption{Yearly MAE (MSCM) on test sets.}
	\label{mae}
\end{table}

\begin{figure}[H]
	\centering
	\includegraphics[width=\textwidth]{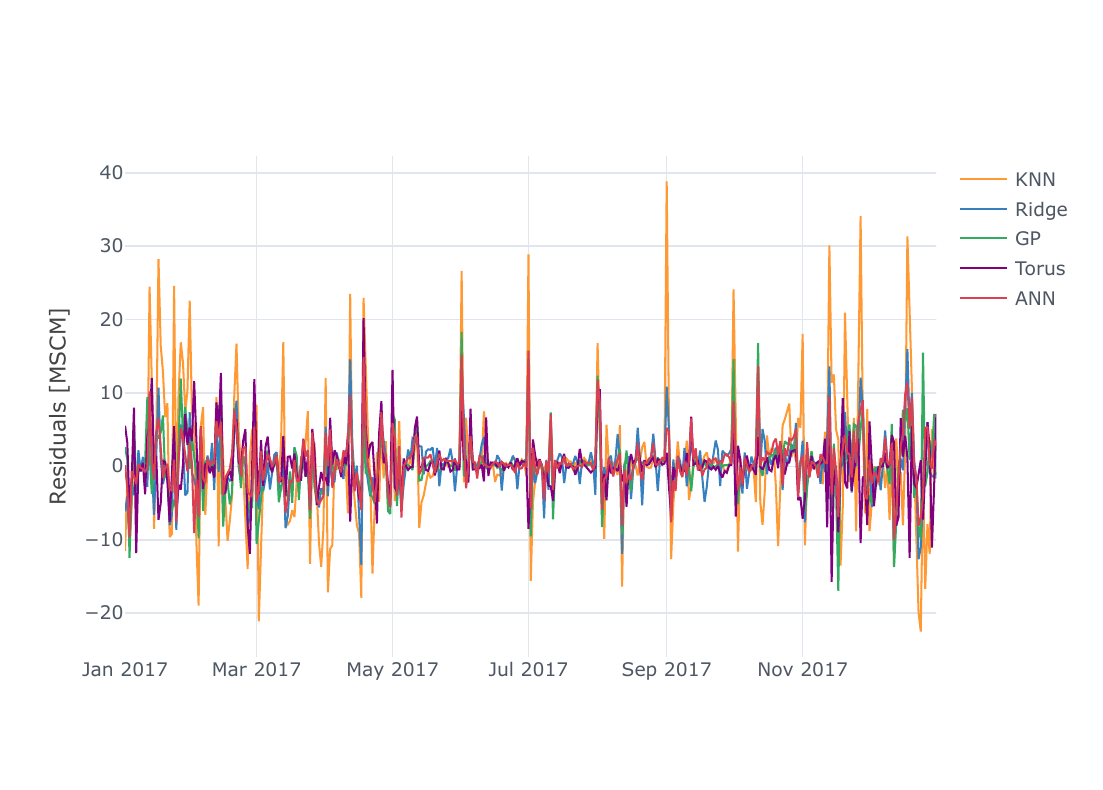}
	\caption{Out-of-sample model residuals in 2017}
	\label{err17_line}
\end{figure}

\begin{figure}[H]
	\centering
	\includegraphics[width=\textwidth]{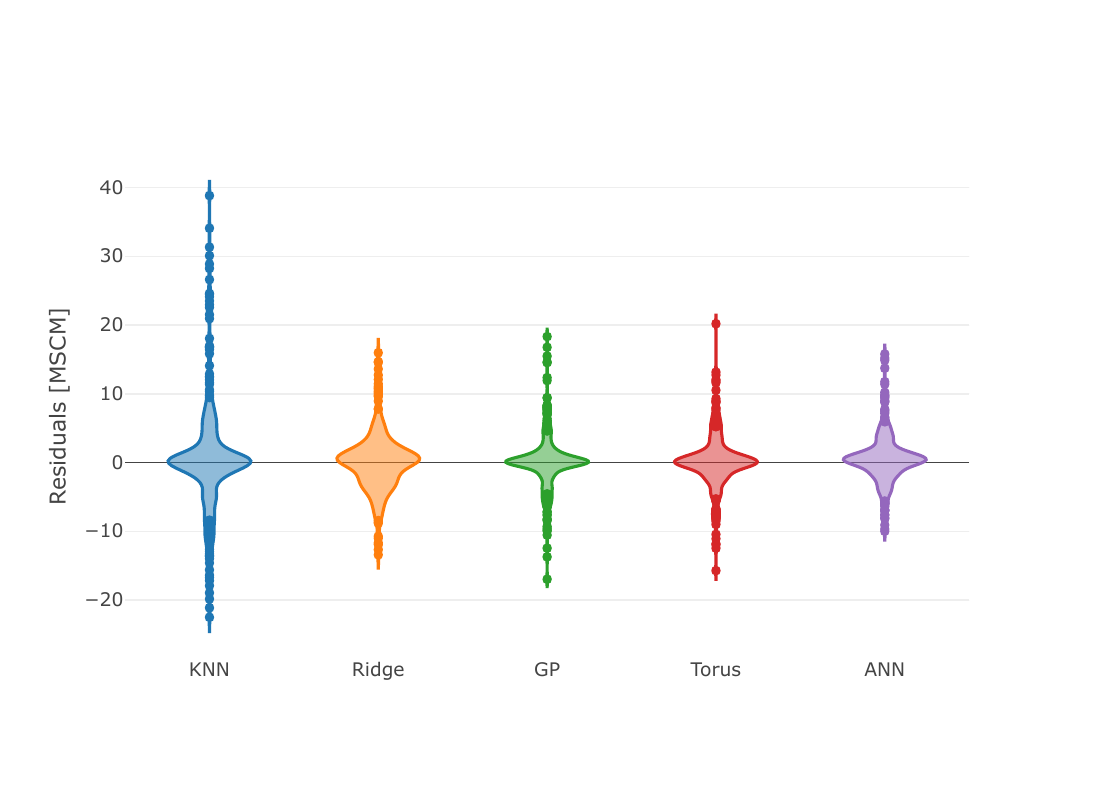}
	\caption{Distribution of out-of-sample residuals in 2017}
	\label{err17_distribution}
\end{figure}

The differences between the RMSE- and MAE-based rankings are possibly explained by the non-Gaussianity of the prediction errors. 
In case of zero-mean prediction errors that are perfectly Gaussian, it should be $\mathrm{MAE}/\mathrm{RMSE}=\sqrt{2/\pi}\sim 0.798$, yielding identical rankings, irrespective of the adopted metrics. As a matter of fact, $\mathrm{MAE}/\mathrm{RMSE}< 0.798$ for all the models: the ratio MAE/RMSE is about 0.61 for GP and Torus, 0.62 for KNN, 0.65 for ANN and 0.72 for Ridge. This is explained by the non-Gaussianity of the prediction errors, possibly associated with the presence of "fat tails" in their distributions. In particular, from Fig. \ref{err17_line} it is apparent that different error variances are observed in the cold and warm seasons. This means that the overall error distribution is akin to a mixture, which can produce fat tails when the variances in the two seasons are much different. The error distributions in 2017 are displayed in Fig. \ref{err17_distribution}. 

Due to the seasonal behavior of RGD, it is of interest to disaggregate data at a monthly level. In \cref{mae_monthly}, the monthly averages of MAE and MAPE are reported throughout the 2015-2017 test years. It appears that GP is the best performer during the warm period, especially from June to October, whereas in the cold months, from December to February, ANN is more accurate. A possible explanation is that the GP model is better at capturing the effects of the weekly seasonality, that explains most of the Summer variability, while ANN accounts better for the non-linear effect of temperature, mostly relevant during the cold months.

To the best of our knowledge, there are no published benchmarks for the forecasting task addressed in this paper. A somehow similar problem was studied by Zhu et al. \cite{zhu2015short} relative to UK gas demand in 2012. Still, their results are not entirely comparable to ours, for two main reasons: first, the authors considered the total UK demand and not just the residential one; second, UK climate is colder than the Italian one.
Nonetheless, we can use a relative error metrics, such as the MAPE, in order to obtain a first level comparison, limited to 6 cold months (from October to March). Our best model in terms of average MAPE over 2015-2017, i.e., the GP, achieves 3.11\%, while Zhu's false neighbors filtered-support vector regression local predictor (FNF-SVRLP) achieves 3.88\% on the same six cold months of 2012. Although no definite conclusion can be drawn, these numbers suggest some degree of consistency between forecasting performances at country level.

\begin{table}[!htbp]
	\begin{center}
	    \resizebox{\columnwidth}{!}{%
		\begin{tabular}{@{}lccccc||ccccc@{}} 
			\toprule
			     &  &  & MAPE(\%) &  &  &  &  & MAE (MSCM) &  &    \\ 
		     Month    & Ridge & GP & KNN & Torus & ANN & Ridge & GP & KNN & Torus & ANN  \\
			\midrule \addlinespace
            January	 &  3.10 &  3.01 &  6.45 &  3.21 &  2.93 &  5.79 &  5.67 &  12.44 &  5.96 &  \textbf{5.45} \\ 
            February	 &  2.75 &  2.84 &  4.77 &  3.33 &  2.62 &  4.52 &  4.59 &  7.60 &  5.48 &  \textbf{4.33} \\ 
            March	 &  3.94 &  4.20 &  7.13 &  3.83 &  4.27 &  4.59 &  4.89 &  8.01 &  \textbf{4.56} &  4.90 \\ 
            April	 &  6.17 &  4.80 &  14.79 &  4.89 &  5.09 &  3.56 &  \textbf{2.89} &  8.23 &  2.98 &  3.04 \\ 
            May	 &  5.76 &  2.67 &  6.08 &  3.21 &  2.50 &  2.37 &  1.22 &  2.67 &  1.51 &  \textbf{1.14} \\ 
            June	 &  4.57 &  1.32 &  6.02 &  3.37 &  1.92 &  1.46 &  \textbf{0.43} &  1.92 &  1.11 &  0.62 \\ 
            July	 &  3.78 &  1.16 &  3.65 &  1.50 &  1.54 &  1.11 &  \textbf{0.35} &  1.13 &  0.45 &  0.46 \\ 
            August	 &  9.39 &  3.00 &  19.44 &  3.86 &  4.50 &  2.24 &  \textbf{0.71} &  4.56 &  0.92 &  1.06 \\ 
            September	 &  5.36 &  1.06 &  3.18 &  1.33 &  1.81 &  1.92 &  \textbf{0.38} &  1.17 &  0.49 &  0.68 \\ 
            October	 &  4.10 &  2.81 &  6.22 &  3.30 &  3.42 &  2.23 &  \textbf{1.78} &  3.91 &  1.99 &  2.09 \\ 
            November	 &  2.70 &  3.14 &  5.50 &  3.03 &  2.9 &  \textbf{3.39} &  3.76 &  6.47 &  3.57 &  3.41 \\ 
            December	 &  2.78 &  2.68 &  4.27 &  2.70 &  2.52 &  4.83 &  4.58 &  7.14 &  4.62 &  \textbf{4.32} \\  
			\bottomrule
		\end{tabular}%
		}
	\end{center}
	\caption{Monthly MAPE and MAE (MSCM) on test sets 2015-2017: best performers in terms of MAE are highlighted in boldface.}
	\label{mae_monthly}
\end{table}

\section{Conclusions} \label{conclusion}

In this paper, one-day-ahead forecasting of the residential gas demand was addressed at the country level. Five different models were considered: Ridge regression, Gaussian Process, K-nearest neighbor, Artificial Neural Network, and the Torus model. The choice of the relevant covariates and the most important aspects of the preprocessing and feature extraction have been discussed, lending particular attention to the role of one-day-ahead temperature forecasts. In particular, a simple model describing the propagation of temperature errors to gas forecasting errors was derived.

The proposed methodology was tested on daily Italian gas demand data from 2007 to 2017. Although a specific benchmark is not available, a comparison with UK data restricted to cold months showed a substantial consistency between the performances achieved in the two countries.

Our best model, in terms of RMSE, was the Artificial Neural network, closely followed by the Gaussian Process. If the MAE is taken as an error measure, the GP became the best model, although by a narrow margin. From the analysis of monthly performance, GP was found to be more accurate in tracking the weekly periodicity, which is predominant in the summer period, while the ANN accounted better for the non-linear influence of temperature, whose contribution is more significant during the winter period. 

An interesting question is how much of the forecasting mean squared error is ascribable to temperature forecasting errors. On the Italian data, we found that the MSE for the ANN model passed from $\mathrm{MSE}=3.65^2=13.32$ (using true temperatures, see \cref{models_with_temp_cons}) to $4.04^2=16.32$ (using temperature forecasts, see \cref{models_vs_perf_limit}). This means that temperature forecast errors account for some 18\% of the MSE of RGD forecasts. As demonstrated in \cref{actual_vs_simulated_rmse}, our error propagation model successfully predicted the quantitative impact of temperature forecast errors on gas forecast errors, a capability that could prove useful in order to assess the extent and convenience of improvement margins associated with more sophisticated (and possibly more expensive) weather forecasts.

Future developments may span along two main directions: first, the study and assessment of more complex architecture of neural networks, such as Recurrent Neural Networks (RNN) and Long-Short Term Memory (LSTM), which have already been successfully applied to power demand forecasting; second, a further refinement of the propagation model of temperature forecasting errors, to remove the restrictive hypothesis of piecewise linear relation between gas demand and temperature.

\bibliographystyle{plain}
\bibliography{references}

\end{document}